\documentclass[preprint2]{aastex}

\usepackage{txfonts,amssymb,subfigure,epsfig}
\usepackage[draft]{hyperref}
\usepackage{multirow}
\usepackage{color}
\usepackage[table]{xcolor}
\usepackage{placeins}
\usepackage{afterpage}

\usepackage{natbib}
\usepackage{natbib}
\newcommand{\gsim}{\lower.7ex\hbox{$\;\stackrel{\textstyle>}{\sim}\;$}}
\newcommand{\lsim}{\lower.7ex\hbox{$\;\stackrel{\textstyle<}{\sim}\;$}}

\newcommand{\PX}{Planet Nine\,}
\newcommand{\ssim}{\,{\sim}\,} 

\slugcomment{Submitted to ApJ XXXX}

\shorttitle{Constraining \PX}
\shortauthors{Holman et al. 2016}

\begin{document}

\title{Observational Constraints on \PX: Astrometry of Pluto and Other Trans-Neptunian Objects}

\author{Matthew~J.~Holman and Matthew~J.~Payne}
\affil{Harvard-Smithsonian Center for Astrophysics, 60 Garden St., MS 51, Cambridge, MA 02138, USA}
\email{mholman@cfa.harvard.edu, mpayne@cfa.harvard.edu, matthewjohnpayne@gmail.com}


\begin{abstract}
We use astrometry of Pluto and other TNOs to constrain the sky location, distance, and mass of the possible additional planet (\PX) hypothesized by \citet{Batygin.2016}.  We find that over broad regions of the sky the inclusion of a massive, distant planet degrades the fits to the observations.  However, in other regions, the fits are significantly improved by the addition of such a planet.  Our best fits suggest a planet that is either more massive or closer than argued for by \citet{Batygin.2016} based on the orbital distribution of distant trans-neptunian objects (or by \citet{Fienga.2016} based on range measured to the Cassini spacecraft).  
The trend to favor larger and closer perturbing planets is driven by the residuals to the astrometry of Pluto, remeasured from photographic plates using modern stellar catalogs~\citep{Buie.2015}, which show a clear trend in declination, over the course of two decades, that drive a preference for large perturbations.
Although this trend may be the result of systematic errors of unknown origin in the observations, a possible resolution is that the declination trend may be due to perturbations from a body, additional to \PX, that is closer to Pluto, but less massive than, \PX.
\end{abstract}


\section{Introduction}\label{SECN:INTRO}

Uranus had made nearly one full orbit since its discovery, when astronomers reported large deviations in its observed sky positions compared with those predicted by available ephemerides~\citep{Bouvard.1824}. Based on those residuals, Le Verrier and Adams~\citep{LeVerrier.1846a,LeVerrier.1846b,Adams.1846} predicted the existence of Neptune, which was discovered shortly thereafter~\citep{Galle.1846}.  That celebrated success ensured that scientists would sporadically revisit the possibility of yet undiscovered planets in the solar system (See \citealt{Tremaine.1990}, \citealt{Hogg.1991} and \citealt{Gaudi.2005} for reviews).

Although similar anomalies in the positions of Neptune motivated the search that ultimately resulted in the discovery of Pluto~\citep{Lowell.1915}, Pluto was not sufficiently massive to account for them.  Improved planetary ephemerides, based on more accurate planetary masses and careful vetting of observational data, eliminated most of those residuals~\citep{Standish.1993,Quinlan.1993}.  

Nevertheless, the discovery of the Kuiper belt~\citep{Jewitt.1993}, including its intricate dynamical structure \citep{Malhotra.1995,Gladman.2012}, detailed size distribution~\citep{Gladman.2001,Bernstein.2004,Fraser.2014}, numerous large members~\citep{Brown.2004,Brown.2005,Schwamb.2010}, and distant components~\citep{Gladman.2002,Brown.2004,Trujillo.2014}, rekindled enthusiasm for the possibility of additional planets, orbiting undetected in the far-reaches of the outer Solar System.  A number of current solar system formation models feature large-scale planet migration~\citep{Fernandez.1996,Malhotra.1993,Malhotra.1995,Levison.2003,Morbidelli.2005,Tsiganis.2005,Levison.2008}, additional giant planets formed $\sim5-35$~AU from the Sun and then scattered within or ejected from the solar system~\citep{Chiang.2007,Bromley.2014,Bromley.2016}, or the formation of planets at larger distances~\citep{Kenyon.2015,Kenyon.2016}.   

An additional planet could leave observable signatures in the orbital distribution of trans-neptunian objects (TNOs).
\citet{Brunini.2002} investigated the effect of a relatively close ($\sim60\,AU$), Mars-mass object embedded in the Kuiper belt (KB) region, proposing that such an object could naturally explain the observed ``edge'' in the KB distribution at $\sim50\,AU$, although later work found that such a body would be inconsistent with other features of the observed orbital distribution of TNOs~\cite{Melita.2004}.  \citet{Lykawka.2008} also simulated the dynamical evolution of the outer solar system under the influence of a planet with a mass a few tenths that of the Earth. They hypothesized that such a body was scattered out by one of the giant planets, causing it to excite the primordial Kuiper belt to the levels observed at 40 - 50 AU while also truncating it at $\sim50\,AU$. Subsequent interactions then pushed the outer planet into a distant ($\gsim 100$ AU) and inclined ($ 20\arcdeg - 40\arcdeg$) orbit.

\citet{Gomes.2006} examined the long-term, secular influence of a planet in the inner Oort cloud on the orbits of scattered disk objects with initial perihelion values near Neptune ($32<q<38$~AU).  They find that such a planet, with a $M_c/b_c^3 > 0.8\times 10^{-14} M_\odot AU^{-3}$, where $b_c=\sqrt{a_c(1-e_c^2)}$ ($a_c$ and $e_c$ being the respective semi-major axis and eccentricity of the companion), can lift the perihelia of scattered-disk objects (SDOs) to $q>75$~AU (values seen for Sedna and 2001~CR105 and other extreme scattered disk objects)  on Gyr time scales.

In their paper announcing the discovery of  2012~VP113, \citet{Trujillo.2014} pointed out that the known ``extreme scattered disk objects'' (semi-major axes $a>150\,AU$ and perihelion distances $q>30\,AU$) have arguments of perihelion clustered near $\omega\approx 340\arcdeg \pm 55\arcdeg$.  They emphasized that observational bias cannot account for this clustering.  Their simulations showed that a super-Earth-mass body at $250\,AU$ can maintain the values of $\omega$ near zero for billions of years.  \citet{Trujillo.2014} speculated that such a planet is responsible for the observed argument of perihelion clustering and note that such a planet, with very low albedo, would be fainter than the detection limits of current surveys. \citet{Fuente_Marcos.2014} used simulations of observational surveys to confirm that the orbital clustering of extreme scattered disk objects pointed out by ~\citet{Trujillo.2014} is unlikely to be the result of observational bias.  

\citet{Gomes.2015} examined the orbital distribution of distant centaurs ($a>250\,AU$) and found that the large fraction of such distant objects which are luminous (compared to classical centaurs), can be best explained by the existence of one or more distant planets in the extended scattered disk.

\citet{Batygin.2016} discovered that the orbits of distant KBOs cluster not only in argument of perihelion, but also in physical space. They found that the observed orbital alignment could be maintained by a distant, eccentric planet of mass $\sim10\,M_{\oplus}$ orbiting in approximately the same plane as the KBOs, but with a perihelion $180\arcdeg$ away. 
In addition they found that such a planet could explain the orbits of high semi-major axis objects with inclinations between $60\arcdeg$ and $150\arcdeg$, whose origin was previously unclear.  Despite its otherwise impressive specificity, the long-term dynamical analysis of \citet{Batygin.2016} does not directly constrain the location of \PX\,within its orbit.

Most of the constraints on the mass and orbit of \PX have come from the observed orbital distribution of TNOs and how \PX could account for peculiar features of that distribution; upper limits from dynamical effects that are not observed \citep{Hogg.1991,Iorio.2009,Iorio.2012,Iorio.2014}; non-detections in optical \citep{Brown.2015} and infrared surveys \citep{Luhman.2014}; and models of the physical properties such a planet might have~\citep{Ginzburg.2016}.
Other investigations have used the simulation of orbital distributions and survey results to constrain the sky plane location of \PX~\citep{Brown.2016, Fuentes_Marcos.2016}.  

However, the analysis of the precision ranging measurements to the Cassini spacecraft \citep{Fienga.2016}, or studies of the resonant interactions that distant objects would have with \PX \citep{Malhotra.2016}, can also favor or rule out particular sky-plane positions for the planet, on a physical basis.
To our knowledge, the Cassini range measurements are the only {\it direct} observations that support the existence, mass, and orbit of \PX.

In the present investigation, we examine the astrometry of Pluto and other TNOs to search for direct evidence of \PX.
Precise positional observations of these bodies have the potential to be particularly sensitive to the mass and orbit of \PX, for a number of reasons.  The astrometry of Pluto spans more than a century (pre-discovery observations date to 1914), and a number of the large TNOs have been observed for several decades.   These long time spans allow the effects of weak perturbations to accumulate.  The large heliocentric distances of these bodies also make them sensitive probes to tidal perturbations.  Finally, the availability of improved stellar catalogs, with reliable proper motions, permit the accurate re-measurement of archival data.

We organize the remainder of this paper as follows.  In section~\ref{SECN:METHOD}, we present our model and numerical methods. In section~\ref{SECN:OBS_DATA}, we describe the observational data to which we apply our model.  In section~\ref{SECN:TILING}, we extend our method to the full sky, and in section~\ref{SECN:RESULTS}, we present those results.  
In \ref{SECN:DISC}, we discuss our conclusions and review the overall constraints on the existence of \PX.


\section{Numerical Method}
\label{SECN:METHOD}

Following the approach of \citet{Hogg.1991}, we test whether fits to the orbits of Pluto and a set of well-observed TNOs are
improved by the inclusion of a massive, distant perturber.
We use $\chi^2$ to evaluate the fit, where $\chi^2$ is approximately
\begin{equation}
    \chi^2 = \sum_i \cos^2(\delta_{o,i}) \left(\frac{\alpha_{o,i}-\alpha_{c,i}}{\sigma_\alpha}\right)^2 + \left(\frac{\delta_{o,i}-\delta_{c,i}}{\sigma_\delta}\right)^2.
\end{equation}
The {\it observed} RA and Dec of the $i$-th observation are $\alpha_{o,i}$ and $\delta_{o,i}$, and the corresponding {\it calculated} values  from the model are  $\alpha_{c,i}$ and $\delta_{c,i}$.  Their respective uncertainties are $\sigma_\alpha$ and $\sigma_\delta$.   
(Internally, our code uses a local tangent plane to evaluate each difference between observed and calculated values, to avoid dependence on a particular coordinate system.)

First, we determine the minimum chi-squared, $\chi_{\mathrm{ref}}^2$, for a reference model that includes the Sun and known planets, as well as the large TNOs as gravitational perturbers.
Then we determine the corresponding value, $\chi_{\mathrm{pert}}^2$, for a model that includes an additional perturbing planet
(Details of the orbit of the perturbing planet are given below).
If the value of $\Delta\chi^2 = \chi_{\mathrm{pert}}^2 - \chi_{\mathrm{ref}}^2 $ is \emph{negative} and
significant, i.e. the fit is improved, we take that as evidence supporting the hypothesis.  
If the $\Delta\chi^2$ is \emph{positive} and sufficiently large, we take that as excluding a
planet with those parameters.

\subsection{Orbit fitting}
We carry out the fits with an extensively modified version of {\it Orbfit} \citep{Bernstein.2000}, software developed to fit the orbits of TNOs to astrometric observations.  
{\it Orbfit} determines the cartesian position and velocity of the fitted TNO in an inertial frame at a given epoch. 
The trajectory of the TNO is integrated in the gravitational field of the Sun and planets to the time of each observation, with   
the topocentric position of the observatory and the light travel time properly accounted for.  The $\chi^2$ of the fit of the model to the observations is minimized  using the Levenberg-Marquardt algorithm~\citep{Press.1986}.  The parameter basis of {\it Orbfit} is orthogonal and provides an approximation to the orbital motion that is nearly linear in the parameters.  Thus, the Levenberg-Marquardt algorithm is well-behaved, in this case.

Our modifications to {\it Orbfit} include the following:
\begin{enumerate}
\item We eliminated {\it Orbfit}'s dependence on the tangent plane approximation.
Although this approximation is well suited to fitting TNO observations that
cover a small number of years, Pluto has traversed nearly $180\arcdeg$ in true
anomaly over the span of its observations.  Thus, a tangent plane does not
accurately capture Pluto's observed trajectory.   
Internally we represent the sky position of the fitted object with a 3-dimensional, topocentric unit vector directed from the observatory to the observed location.  However, when comparing the calculated sky position with those observed, we resolve the differences along the RA and Dec directions using local tangent plane projections~\citep{Herget.1965}.

\item We replaced {\it Orbfit}'s analytic but approximate expressions for the derivatives
of the observable properties as a function of orbital parameters
(required by the Levenberg-Marquardt minimization routine) with numerical but fully accurate derivatives.  

\item We included the Sun and all of the known planets as perturbers, rather than just the four giant planets. 
In addition, we include 10 massive TNOs as perturbers: see Secn. \ref{SECN:LARGE_TNOS} for further discussion of the perturbation magnitudes arising from the known TNOs.

\item We upgraded the underlying ephemeris from JPL's DE405 to
DE432.

\item We replaced {\it Orbfit}'s calculation of the
topocentric position of the observatory with routines that use current
data for the geocentric observatory positions from the MPC.  
\item We reduced the time step in {\it Orbfit}'s symplectic 
leap-frog integrator to 10 days from 20 days.  
\end{enumerate}
We emphasize that these modifications are only necessary because Pluto has been observed for
such a long orbital arc and because we are want to ensure that our
results are not limited by the accuracy of the model.  {\it Orbfit} results for typical TNOs investigations are still valid.
\footnote{We will describe these modifications in greater detail in a forthcoming publication.  The code is available upon request.}

\subsection{Tidal perturbations from \PX}
The principal effect of a distant planet is to accelerate all of
the masses in the solar system toward that planet.   Any perturbation that leaves the relative positions and velocities of the Sun
and planets essentially unchanged, and that does not significantly accelerate
the solar system barycenter, would be unobservable
(The uniform velocity of the solar system barycenter is removed in the DE432 ephemeris, as is the case for other JPL ephemerides).
Only changes in the relative position or velocity between an observing station and some
other reference position or velocity can be detected.  Thus, in our case, only the effects of tidal
acceleration between the observing station and the observed object are important.

As pointed out by \citet{Tremaine.1990} and \citet{Fienga.2016}, the most conservative approach is to vary all of the parameters for all of the bodies when fitting an ephemeris model to the observations.  
However, to simplify our analysis, we assume that Sun and planets follow the fixed barycentric trajectories given by the DE432 ephemeris and are not affected by \PX, and we vary only the barycentric orbital parameters of Pluto and other TNOs.  
We have three justifications for this approach:
First, the DE432 ephemeris
is an accurate fit of a very comprehensive dynamical model to essentially all of the
relevant, constraining observations of the solar system bodies.  There is
little room for the Sun and planets to deviate from the trajectories specified by DE432 while satisfying the observational constraints; 
Second, the planets only weakly perturb the orbits of Pluto and other TNOs.
Small changes in the trajectories of the solar system planets themselves due to an additional planet will lead to negligible changes to the perturbations of the solar system planets on Pluto and the other TNOs; 
Third, changes in the location of the observing sites (primarily through changes in the location of the geocenter), relative to the solar system barycenter, would be a factor of $\sim 30$ smaller than corresponding changes in the location of Pluto or other TNOs due to any additional acceleration (see below).  
In Fig \ref{FIG:RESID2}, we show the results of full n-body integrations from initial conditions of the Sun, planets, \PX, and Pluto.  We obtain the  initial conditions at a reference time from our {\it Orbfit} modeling.  In the figure we show the difference between the sky position of Pluto when \PX is included and when it is not.  This demonstrates that the results of a model in which the positions of the planets are allowed to vary are essentially indistinguishable from our standard model, supporting our simplified model.

We include the acceleration of Pluto and TNOs due to an
additional planet in two different ways.  First, we explicitly incorporate the relative acceleration between Pluto (or the other TNOs) and the barycenter due to \PX into the equations of motion.  The additional acceleration of Pluto is given by
\begin{equation}
    \ddot{\vec{a_P}} = -\frac{GM_X}{r_{PX}^3}\left( \vec{r}_P-\vec{r}_X\right) - \frac{GM_X}{r_{X}^3}\vec{r}_X,
    \label{EQN:MOVING}
\end{equation}
where $\vec{r}_P$ is the barycentric position of Pluto (or TNO), $M_X$ and $\vec{r}_X$ are the respective mass and barycentric position vector of the distant planet, $r_P=|\vec{r}_P|$, $r_X=|\vec{r}_X|$, and $r_{PX}=|\vec{r}_P-\vec{r}_X|$.  The primary approximation, as noted above, is that the additional planet accelerates the barycenter of the solar system, rather than the individual solar system bodies.  We further assume that \PX follows a keplerian orbit and is itself unperturbed by the other planets.   We refer to this approach as the ``moving planet model.''

For our second approach, following \citet{Hogg.1991}, we assume that the additional planet is stationary and is sufficiently distant that its gravitational potential can be approximated as 
\begin{equation}
    \Phi(r_P) = \frac{G M_X}{2 {r_X}^3}\left[  r_P^2 - 3\left(\vec{r}_P\cdot{\hat r }_X \right)^2 \right]
    \label{EQN:TIDE}
\end{equation}
where ${\hat r}_X=\vec{ r }_X/r_X$, and 
the corresponding tidal acceleration is given by
\begin{equation}
    \ddot{\vec{a_P}} = -\frac{GM_X}{r_{PX}^3}\left[\vec{r}_P - 3\left(\vec{r}_P \cdot {\hat{r}}_X\right) {\hat{r}}_X\right].
\end{equation}
The strength of the tidal perturbation scales with $M_X/r_{PX}^3$.  We refer to this approach as the ``tidal model''.
We note that if \PX has a semi-major axis $\sim700\,AU$ \citep{Batygin.2016}, its period will be $\sim18,000\,$ years, and hence over the $\sim85\,$ year span of our observations, \PX is approximately stationary on the sky. 

\subsection{Large TNOs}
\label{SECN:LARGE_TNOS}
%
\begin{table*}[t]
\begin{minipage}[th]{\textwidth}
\begin{center}
\begin{tabular}{ |cl|c|c|c|c| } 
 \hline
          &           & Separation    & Mass         &  $M/r^3$ &                 Mass    \\
Designation & Name    & ($AU$)        & ($M_\sun$)   & ($M_{\odot\,AU^{-3}}$) & Reference \\
 \hline
(136199)&Eris         & 102-128 & $8.35\times10^{-9}$ &   $3.3\times10^{-16} - 6.6\times10^{-16}$ & \citet{Brown.2007}\\ 
(134340)&Pluto+Charon & --      & $7.33\times10^{-9}$ &   -                                       & \citet{Stern.2015} \\ 
(136108)&Haumea       & 11-58   & $2.01\times10^{-9}$ &   $3.4\times10^{-15} - 5.3\times10^{-13}$ & \citet{Ragozzine.2009} \\
(50000) &Quaoar       & 13-39   & $7.0 \times10^{-10}$&   $1.2\times10^{-14} - 3.2\times10^{-13}$ & \citet{Fraser.2013}\\
(90377) &Sedna        & 108-130 & $5.0\times10^{-10}$ &   $3.2\times10^{-16} - 5.6\times10^{-16}$ & \emph{Assumed}   \\ 
(90482) &Orcus        & 35-73   & $3.2\times10^{-10}$ &   $1.8\times10^{-15} - 1.6\times10^{-14}$ & \citet{Carry.2011} \\
(307261)&2002 MS4     & 17-55   & $3.2\times10^{-10}$ &   $4.2\times10^{-15} - 2.1\times10^{-13}$ & \emph{Assumed}   \\ 
(120347)&Salacia      & 47-71   & $2.2\times10^{-10}$ &   $2.0\times10^{-15} - 6.7\times10^{-15}$ & \citet{Stansberry.2012} \\
(136472)&Makemake     & 16-68   & $1.76\times10^{-10}$ &  $2.2\times10^{-15} - 1.7\times10^{-13}$ & \emph{Assumed}   \\ 
(225088)&2007~OR10    & 68-113  & $1.76\times10^{-10}$ &  $4.9\times10^{-16} - 2.2\times10^{-15}$ & \emph{Assumed}   \\ 
 \hline
\end{tabular}
\caption{The TNOs which were assigned as massive perturbers within our version of {\it Orbfit}. The ``Separation'' refers to the range of distances between the body and Pluto over the course of the observation set for Pluto. ``\emph{Assumed}'' values for TNO masses were calculated using the TNO (of known mass) with the nearest mean diameter, and assuming mass scales with diameter cubed.}
\label{TABLE:TNOs}
\end{center}
\end{minipage}
\end{table*}
There are a number of large TNOs that might also perturb the orbits of Pluto and other TNOs.  Although these are all substantially less massive than \PX might be, they are much closer.  Thus, in addition to the gravitational perturbations from the Sun and planets, we include the perturbations from a set of large TNOs.  The masses of the majority are known from the orbits of their satellites.  For the rest we estimated masses. The large TNOs and their properties are provided in Table \ref{TABLE:TNOs}. Paralleling our assumption that the Sun and planets follow the trajectories given by DE432, we assume that the large TNOs follow fixed orbits.


Over the span of Pluto's observations, the acceleration of Pluto due to other TNOs at a given time can exceed that from \PX, for the  smaller tidal parameters of \PX.  For example, the direct acceleration due to Haumea exceeds the tidal acceleration from \PX for several years in our simulations.  This does not imply that the long-term effect of those accelerations is larger than that from \PX.  It is simply a measure of the effect on shortest time scales.  The tidal influence from \PX, with its acceleration oriented in the same direction over long periods, can dominate over other accelerations that tend to average out.  Similarly, resonant interactions between Pluto and other bodies, even those with small masses, can dominate.

\section{Observational Data}
\label{SECN:OBS_DATA}

In order to guide the trajectory of the New Horizons spacecraft to its encounter with the Pluto system~\citep{Stern.2015}, a number of groups concentrated on improving the determination of Pluto's orbit~\citep{Assafin.2010,Benedetti.2014,Girdiuk.2014,Pitjeva.2014,Buie.2015}.

As part of that effort, \citet{Buie.2015} used the DASCH~\citep{Grindlay.2009} and DAMIAN~\citep{Robert.2011} scanning systems to remeasure a collection of photographic plates of Pluto taken by Carl Lampland at Lowell Observatory from 1930 through 1951.  
The primary motivation for the re-analysis was to identify and correct any significant but previously unrecognized systematic errors in the historical astrometry of Pluto that might be affecting its orbit determination.  JPL's DE432 ephemeris incorporates the results of the \citet{Buie.2015} analysis. The 3-dimensional positional uncertainty of the Pluto system just prior to the encounter was $\sim1000$~km. The successful New Horizons flyby of the Pluto system demonstrates the accuracy with which Pluto's location and orbit are currently known.  

Initially, we considered using all of the astrometric positions of Pluto and TNOs 
available from the Minor Planet Center's (MPC).  However, 
the data set is heterogeneous and lacks reported astrometric uncertainties.  Thus,
we decided to build upon the carefully selected data of
\citet{Buie.2015} for our analysis of Pluto.  We use the MPC astrometry for TNOs only.

For Pluto, we include the astrometry from the
remeasured Lampland plates~\citep{Buie.2015}; that from a selection of photographic plates from Pulkovo Observatory that were also remeasured with modern stellar catalogs~\citep{Rylkov.1995}; Pluto or Charon positions from recent occultation measurements of Pluto and Charon~\citep{Assafin.2010,Benedetti.2014}; and CCD observations from Pico dos Dias Observatory~\citep{Benedetti.2014}, the
USNO's Flagstaff Station~\citep{Stone.2003}, and JPL's Table Mountain Observatory (described in \citealt{Buie.2015}).  Note that, like \citet{Buie.2015}, we exclude the Lowell data from 1930, as those data appear to have a systematic trend that may be due to systematic uncertainties in their observation times.

We adopt the uncertainties of \citet{Buie.2015} for the Pluto astrometry from Lowell and Pulkovo Observatory.  However, we used significantly smaller astrometric uncertainties for the Pico do Dias data.  \citet{Buie.2015} argued that the observations taken within a single night would have correlated errors due to the astrometry being measure by a common set of stars.  Accordingly, they increased the reported astrometric uncertainties by a factor $\sqrt{N}$, where $N$ is the number of observation with a night.  We find that this overestimates the uncertainties, based on the post-fit RMS.  We find that increasing the astrometric uncertainties of \citet{Benedetti.2014} by a factor $\sim2$ results in a match with the post-fit RMS.  

We also used somewhat smaller astrometric uncertainties for the remaining data sets.   For the USNO astrometry, we adopt $0\,\farcs09$ for both RA and Dec.
For the Table Mountain Observatory astrometry, we adopt $0\,\farcs07$ and $0\,\farcs05$ for the RA and Dec, respectively.  For the occultation data we adopt $0\,\farcs05$ and $0\,\farcs03$ for the RA and Dec, respectively.
The astrometric uncertainties we adopted result in a $\chi^2$ per degree of freedom $=1$ for the unperturbed Pluto orbit fits.

We included all TNOs, including SDOs, with semi-major axes $a\gsim30$~AU for which we could fit reliable orbits.  
For these objects, we adopt a fixed astrometric uncertain of $0\,\farcs27$, which results in $\chi^2$ per degree of freedom $=1$ for the ensemble of TNOs.

We include a total of 6,677 observations for Pluto, plus 35,646 observations of other TNOs \footnote{We provide online tables of all observations \emph{\bf (Insert link at time of publication)}}.  

Despite the larger number of observations of TNOs other than Pluto, the Pluto observations are more constraining.  As noted earlier, the Pluto observations that we use were collected as part of systematic observing programs~\citet{Buie.2015}.  The astrometric uncertainties are typically significantly smaller and the time span is much greater for the Pluto observations than for those of the other TNOs.  Although a number of the bright TNOs also have archival photographic astrometry  that increase the time spans to greater than 60 years, those observations were serendipitous, rather than part of a systematic observing programs.  In all cases, there are just a few archival data points that are separated from rest by decades.  Although the archival observations tightly constrain the orbital elements of the TNOs, they do not appear to add significantly to the constraints on \PX.  That could change if the plates were re-measured, but we are not aware of a program to systematically re-measure those photographic plates, such as the \citet{Buie.2015} program for the Lampland plates.

\section{Tiling the Sky}
\label{SECN:TILING}

The gravitational influence of a perturbing planet depends upon its direction and distance from other bodies in the solar system.  
We use the HEALPix tiling of the sky \citep{Gorski.2005} to test a complete set of perturbing planet locations, uniformly distributed on the sky.
For the results presented in Section \ref{SECN:RESULTS}, we concentrate on the $N_{\mathrm{side}}=2^4=16$ resolution level, resulting in 3072 tiles.
We interpret each HEALPix location as representing an individual possible perturber position in an equatorial frame. We then transform these within {\it Orbfit} to ecliptic coordinates, as well as the body-specific projection coordinates used in {\it Orbfit} \citep{Bernstein.2000}.
For a given HEALPix tile, i.e a given RA \& Dec, we study a range of distances and masses for \PX.  

When we adopt the tidal potential of a stationary, distant planet, nothing more needs to be specified, as the object is stationary.  

In the case of a moving planet, we assume its orbit is circular ($e=0$) and prograde with semi-major axis $a=r_X$. For components of the unit vector to \PX in ecliptic coordinates $x$, $y$, and $z$, we take the inclination to be $i=\sin^{-1} z$.  We adopt $\omega=\pi/2$ for $z\geq0$ and $\omega=-\pi/2$ for $z<0$ for the argument of perihelion, and $\Omega = \theta - \omega - M_0$ for the longitude of ascending node, where $\theta=\tan^{-1}(y, x)$ is the ecliptic longitude and $M_0$ is the mean anomaly at a reference date, which we take to be zero at JD 2436387.0 (1958 July 2.5).  This places \PX at the specified location toward the middle of our data span for Pluto.

To test the dependence of our results on the rate and direction of motion of \PX, we also explored retrograde and polar orbits, following a similar approach.  As we demonstrate below, our results do not depend sensitively to the details of \PX's orbit.

\section{Results}
\label{SECN:RESULTS}

\begin{figure*}
    \begin{minipage}[b]{\textwidth}
    \centering
    \includegraphics[trim = 0mm 0mm 0mm 0mm, clip, angle=0, width=\columnwidth]{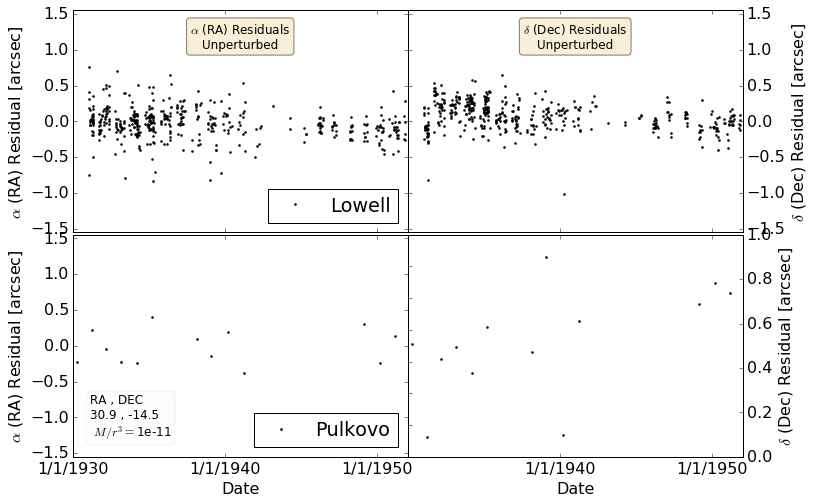}
    \caption{ Zoom-in on the \emph{unperturbed} model of Figure \ref{FIG:RESID1} for the 1930-1950 period, plotting the residuals (observations - model), where the observations taken at different observatories are plotted in different colors. Y-axis scale chosen to facilitate comparison with Buie \& Folkner (2014).
    }
    \label{FIG:RESID1A}
    \end{minipage}
    \end{figure*}

\begin{figure*}
    \begin{minipage}[b]{\textwidth}
    \centering
    \includegraphics[trim = 0mm 0mm 0mm 0mm, clip, angle=0, width=\columnwidth]{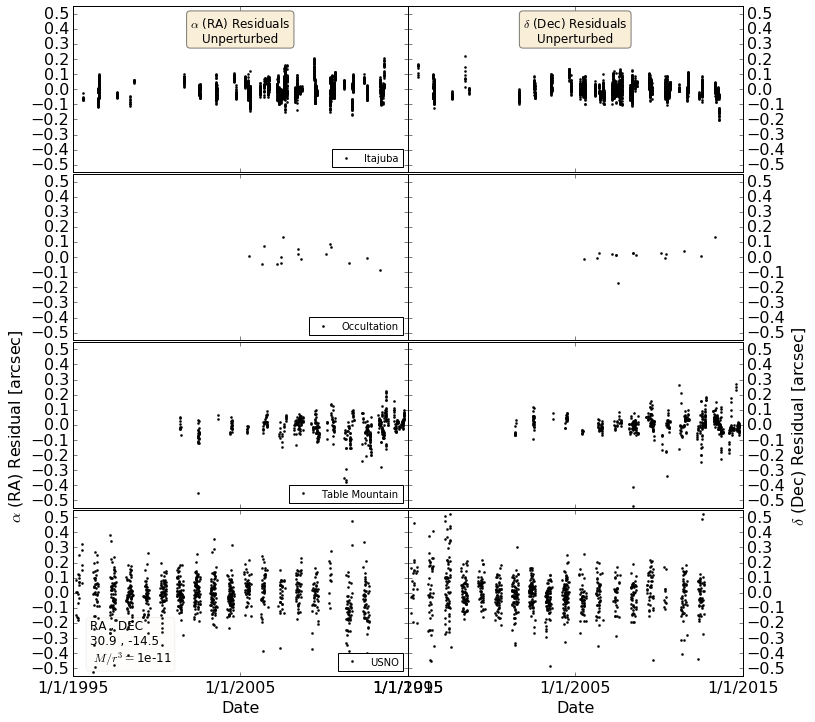}
    \caption{ Zoom-in on the \emph{unperturbed} model of Figure \ref{FIG:RESID1} for the 1970-2010 period, plotting the residuals (observations - model), where the observations taken at different observatories are plotted in different colors. Y-axis scale chosen to facilitate comparison with Folkner et al. (2014). 
    }
    \label{FIG:RESID1B}
    \end{minipage}
    \end{figure*}

Using the methods described in Section \ref{SECN:METHOD}, we perform fiducial orbit fits for Pluto and all the other TNOs in an unperturbed Solar System model (i.e. with \emph{no} \PX).  The resulting orbits of Pluto and the other TNOs are in detailed agreement this those available from JPL or the MPC.  Figures~\ref{FIG:RESID1A} and \ref{FIG:RESID1B}  facilitate comparison with the fits of \citet{Folker.2014} and \citet{Buie.2015} with our results.  

    %
    %

%
\subsection{Level of Significance}
\label{SECN:METHOD:SIG}

After determining the best-fit orbit of Pluto assuming no additional perturber, we now evaluate whether the inclusion of \PX improves or degrades the fits.

As noted in section~\ref{SECN:TILING}, we evaluate our two models, tidal and moving planet, at a full range of locations on the sky.   At each of those locations the direction of the perturbation, either the tidal perturbation vector or the location of \PX at a reference epoch, is fully specified by two position angles.
We add a perturbation due to \PX using the moving planet and tidal methods of Eqn.\ref{EQN:MOVING} and Eqn.\ref{EQN:TIDE}, respectively.

For the tidal model, we introduce one additional degree of freedom: the magnitude of the perturbation (which scales as $M_X\,r_X^{-3}$ as per Equation \ref{EQN:TIDE}).
For the moving planet model, we introduce two additional degrees of freedom:  the separate mass and semi-major axis of the planet. 

If, in reality, there is no additional planet in the solar system, the fits will be slightly improved simply by introducing additional degrees of freedom, $\Delta\chi^2=1$ and $\Delta\chi^2=2$, for tidal and moving planet models, respectively.  However, with $3-\sigma$ confidence, we expect $\Delta\chi^2<9$ and $\Delta\chi^2<11.4$ \citep{Press.1986}.  More substantial improvements in the fits can be taken as ruling out the null hypothesis that there is no additional planet or unmodeled acceleration.  

Likewise, if the inclusion of \PX {\it degrades} the fits by more than these thresholds, we take that as strongly favoring the null hypothesis over an additional planet, i.e. excluding a perturber with those parameters.

We caution that these confidence intervals assume the errors are normally distributed and uncorrelated, which is not the case.   
However, our results do not depend sensitively on the threshold value of $\Delta \chi^2$.

\subsection{Perturbed Orbital Fits}
\label{SECN:RESULTS:TIDAL}

\begin{figure}[p]
    \centering
    \includegraphics[trim = 0mm 0mm 0mm 0mm, clip, angle=0, width=\columnwidth]{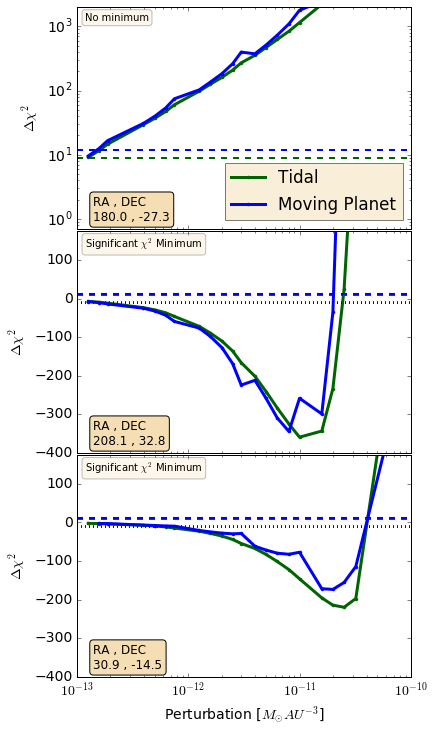}
    \caption{ %
    Examples of $\Delta\chi^2$ for Pluto as a function of perturbation magnitude.
    Results for the tidal perturbation model (eqn.~\ref{EQN:TIDE}) are plotted in \emph{green}.
    Those for the full orbiting planet model (eqn.~\ref{EQN:MOVING}) are shown in \emph{blue}.
    At the top we plot results for a perturber    at $RA,Dec=180\arcdeg,-27\arcdeg$.
    In the middle we plot results for a perturber at $RA,Dec=208\arcdeg,32\arcdeg$.
    On the bottom we plot results for a perturber at $RA,Dec=30.9\arcdeg,-15.4\arcdeg$.
    %
    %
    Dashed horizontal lines indicate the $3\sigma$ confidence limits. Perturbations causing $\Delta\chi^2$ greater than this limit can be excluded.
    For regions of the sky with results similar to those at the top we can only provide \emph{upper limits} on the perturbation scale (dashed lines in the above plots) which are within $\Delta\chi^2_{3\sigma}$ of zero. In this top plot we rule-out perturbations $\gsim2\times10^{-13}\,M_{\odot}\,AU^{-3}$.
    For regions of the sky similar to the middle and bottom plots, not only can we provide upper limits, but we can also find the perturbation scale at the absolute minimum $\Delta\chi^2$.
    }
    \label{FIG:EXAMPLES}
    \end{figure}

We begin by illustrating in Figure \ref{FIG:EXAMPLES} the results we obtain for a perturber at a few specific locations on the sky.

For each of these locations, we define a unit vector toward that point and then explore a variety of masses and semi-major axes, which span a range of $10^{-14}-10^{-10}\,M_{\odot} AU^{-3}$ in the tidal perturbation.
(For reference $M_X\ssim10M_\earth$ at $R_X\ssim600$~AU yields a tidal parameter $M_X/R_X^3\ssim 1.4\times10^{-13}$.
For both the tidal and moving-planet perturbation models we refit all orbits using {\it Orbfit} and calculate the change in  $\Delta\chi^2 = \chi_{pert}^2 - \chi_{ref}^2 $ as described in Section \ref{SECN:METHOD}.


    %
    %

In Figure \ref{FIG:EXAMPLES} we plot $\Delta\chi^2$ as a function of the perturbation magnitude for three different example locations on the sky. 
We find that the tidal model and the full orbiting planet model are similar in all cases.  The small variations that can be seen in the moving planet models are due to the spacing of the grid points in mass and semi-major axis, rather than being a numerical artifact.

In the top panel, we plot results for a perturber at $RA,Dec=180\arcdeg,-27\arcdeg$.  For regions similar to this example, all perturbation magnitudes lead to a \emph{worsening} of the fit, and hence we can only provide \emph{upper limits} on the perturbation scale (dashed lines in the above plots) which are within $\Delta\chi^2_{3\sigma}$ of zero. In this top plot we rule-out perturbations $\gsim2\times10^{-13}\,M_{\odot}\,AU^{-3}$. 

In the middle and bottom panels, we plot results for perturbers at $RA,Dec=208\arcdeg,32\arcdeg$ and at $RA,Dec=30.9\arcdeg,-15.4\arcdeg$, respectively.  For regions of the sky similar to these, some perturbation magnitudes \emph{improve} the fits.  Thus, we can find the perturbation magnitude at the $\Delta\chi^2$ minimum, the range of perturbation magnitudes within $\Delta\chi^2_{3\sigma}$ of that minimum, and an upper limit on the perturbation.  For both of these example regions we see that the best-fit perturbation is approximately $10^{-11} M_{\odot}\,AU^{-3}$.

\begin{figure*}
    \begin{minipage}[b]{\textwidth}
    \centering
    \includegraphics[trim = 0mm 0mm 0mm 0mm, clip, angle=0, width=\textwidth]{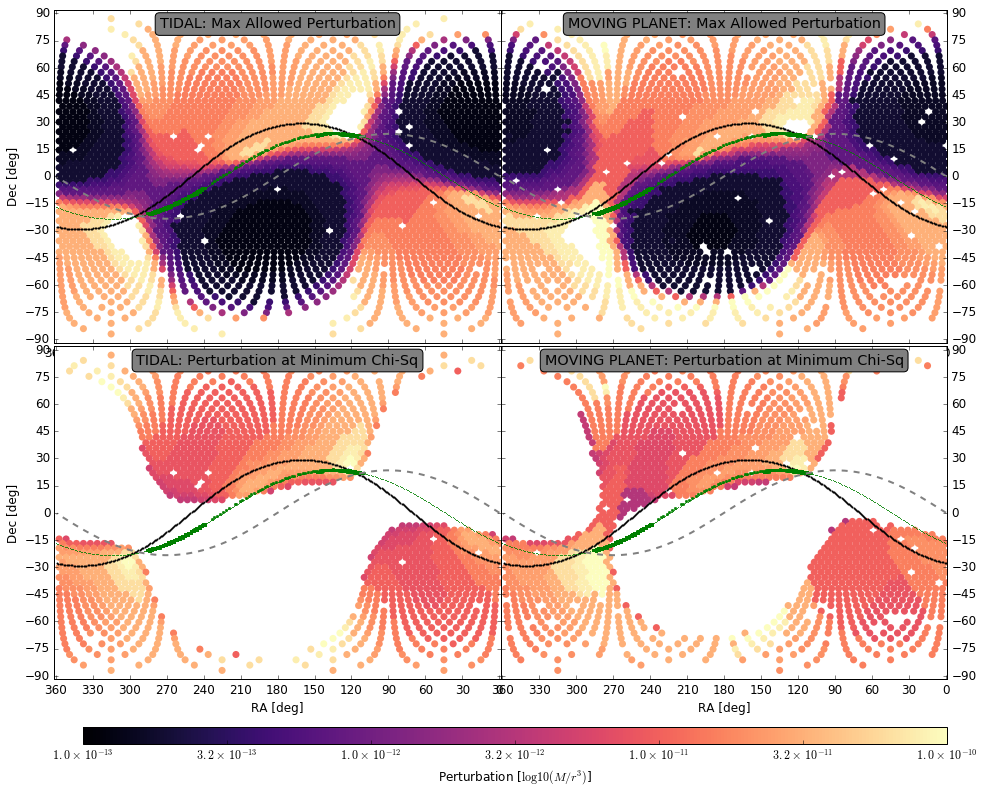}
    \caption{ %
    All-sky constraints on the perturbation from \PX, derived from simultaneous orbital fits to $>500$ TNOs (including Pluto). 
    The plots on the left show the results from the tidal model of Eqn. \ref{EQN:TIDE}, while the plots on the right are from the full moving-planet model of Eqn. \ref{EQN:MOVING}.
    The plots at the top illustrate the maximum allowed perturbation (corresponding to the dashed-lines in Fig. \ref{FIG:EXAMPLES}).
    The plots at the bottom illustrate the perturbation scale at the absolute minima (corresponding to the dotted lines from the bottom plot in Fig. \ref{FIG:EXAMPLES}).
    To guide the eye we plot the ecliptic (gray dashed line), Pluto (green line) and the nominal orbit of \PX (black line) from \citet{Batygin.2016}. %
    Note the orbit of Pluto also has the observational data points over-plotted in larger green points.
    It can be seen that the tidal results on the left are extremely similar to those of the full orbiting planet model across the entire sky.
    }
    \label{FIG:ALLSKY1}
    \end{minipage}
    \end{figure*}

We now divide the  sky into $3,072$ regions using the HEALPix tessellation described in Section \ref{SECN:TILING}, and repeat the fitting procedure illustrated in Figure \ref{FIG:EXAMPLES} at each location.
We plot these all-sky results in Figure \ref{FIG:ALLSKY1}.

In Figure \ref{FIG:ALLSKY1}, the plots on the left show the results from the tidal model of Eqn. \ref{EQN:TIDE}, while the plots on the right are from the full moving-planet model of Eqn. \ref{EQN:MOVING}.
The plots at the top illustrate the maximum allowed perturbation (corresponding to the dashed-lines in Fig. \ref{FIG:EXAMPLES}).
The plots at the bottom illustrate the perturbation scale at the absolute minima.

We draw a few key conclusions from Figure \ref{FIG:ALLSKY1}:
\begin{enumerate}
    \item The results from the tidal and moving planet models are essentially identical.  The figures only differ near the margins.
    \item There are broad regions of the sky (located on the opposite side of the ecliptic from the orbit of Pluto) for which we can rule out perturbation magnitudes $\gsim 3\times10^{-13}\,M_{\odot}\,AU^{-3}$. 
    \item For regions straddling the orbit of Pluto, we can only limit the perturbations to being $\leq\,3\times10^{-11}\,M_{\odot}\,AU^{-3}$.
    Much of this same region also permits a significant decrease in $\chi^2$ of the kind illustrated at the middle and bottom of Figure \ref{FIG:EXAMPLES}, with a broad minimum centered around $\sim\,10^{-11}\,M_{\odot}\,AU^{-3}$. 
    \item We add to the physical and dynamical alignments already noted in the literature \citep{Trujillo.2014,Batygin.2016,Malhotra.2016}, and highlight an interesting (but possibly coincidental) dynamical alignment between the ascending node of Pluto and the proposed ascending node for \PX. 
\end{enumerate}

    %

\begin{figure*}
    \begin{minipage}[b]{\textwidth}
    \centering
    \includegraphics[trim = 0mm 0mm 0mm 0mm, clip, angle=0, width=\columnwidth]{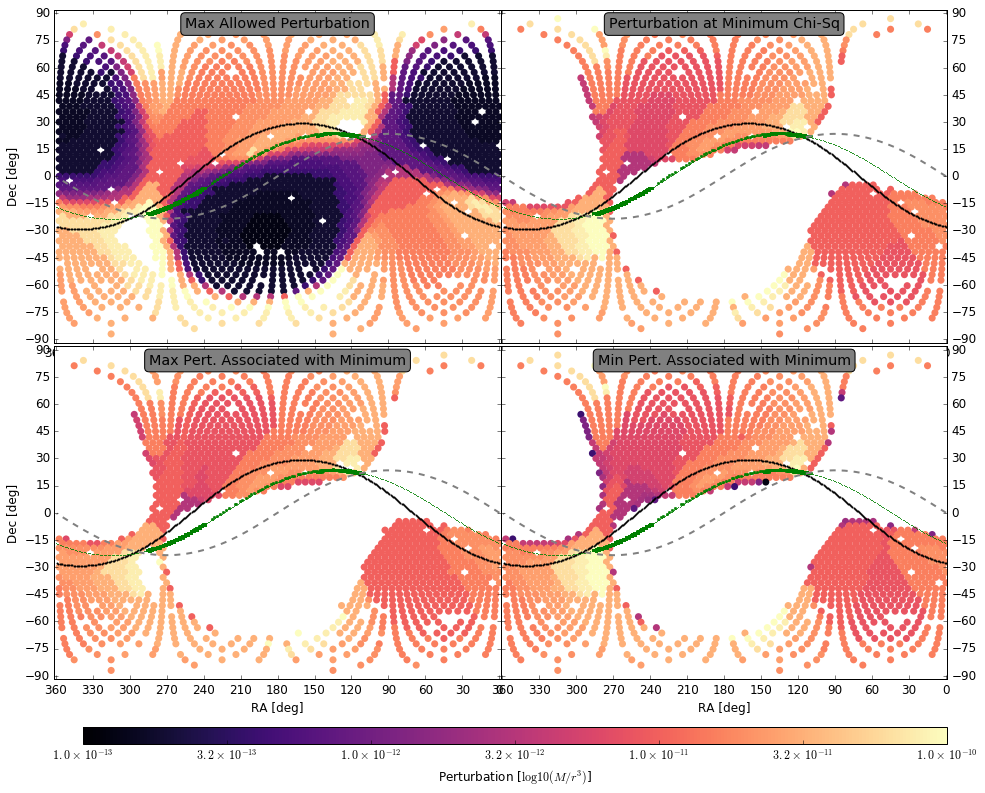}
    \caption{ %
    All-sky constraints on the perturbation from \PX, derived from simultaneous orbital fits to $>500$ TNOs (including Pluto): all plotted results are from the full moving-planet model.
    The plots at the top-left illustrate the maximum allowed perturbation (corresponding to the dashed-lines in Fig. \ref{FIG:EXAMPLES}).
    The plots at the top-right illustrate the perturbation scale at the absolute minima (corresponding to the dotted lines from the bottom plot in Fig. \ref{FIG:EXAMPLES}).
    The plots at the bottom-left illustrate the minimum allowed perturbation within $\Delta\chi^2_{3\sigma}$ of the absolute minimum.
    The plots at the bottom-right illustrate the maximum allowed perturbation within $\Delta\chi^2_{3\sigma}$ of the absolute minimum.
    To guide the eye we plot the ecliptic (gray dashed line), Pluto (green line) and the nominal orbit of \PX (black line) from \citet{Batygin.2016}. 
    Note the orbit of Pluto also has the observational data points over-plotted in larger green points.
    It can be seen that over large regions of the sky we can exclude perturbation scales larger than $\sim 3\,\times 10^{-13} M_{\odot}\,AU^{-3}$ (black and purple regions, top left plot). 
    Close to the orbit of Pluto and the nominal path of \PX, we can only rule out perturbations larger than $\sim 3\,\times 10^{-11} M_{\odot}\,AU^{-3}$
    The plots at the top-right and in the bottom-row illustrate that large perturbations ($\gsim10^{-11}M_{\odot}\,AU^{-3}$:orange regions) are able to significantly improve the fits. 
    }
    \label{FIG:ALLSKY2}
    \end{minipage}
    \end{figure*}

To elaborate on the moving planet model results from Figure \ref{FIG:ALLSKY1}, we show in Figure \ref{FIG:ALLSKY2} the minimum and maximum perturbations which are permissible \emph{while retaining  $\left/\Delta\chi^2\right/ > \Delta\chi^2_{3\sigma}$}.

These plots demonstrate that in the region broadly straddling the orbit of Pluto, the minima are relatively \emph{narrow} for an individual healpix, but difference regions of the sky permit perturbations in the range $10^{-12} - 10^{-10}\,M_{\odot}\,AU^{-3}$.

    %

In Figure \ref{FIG:TRUE_ANOM} we provide further illustration of our constraints, comparing the perturbation magnitudes we favor (and rule out) with the perturbation magnitudes favored (and ruled out) by \citet{Fienga.2016} for the specific nominal orbit of \citet{Batygin.2016}. 
Here it can be seen that 
(a) the constraints placed on the perturbation magnitude of \PX are significantly tighter from the Cassini ranging results of \citet{Fienga.2016}, and 
(b) the unmodelled residuals in the Pluto data demand a very high tidal perturbation to ``correct'' them.
We discuss these issues further in Section \ref{SECN:DISC}.

    \begin{figure*}
    \begin{minipage}[b]{\textwidth}
    \centering
    \includegraphics[trim = 0mm 0mm 0mm 0mm, clip, angle=0, width=\columnwidth]{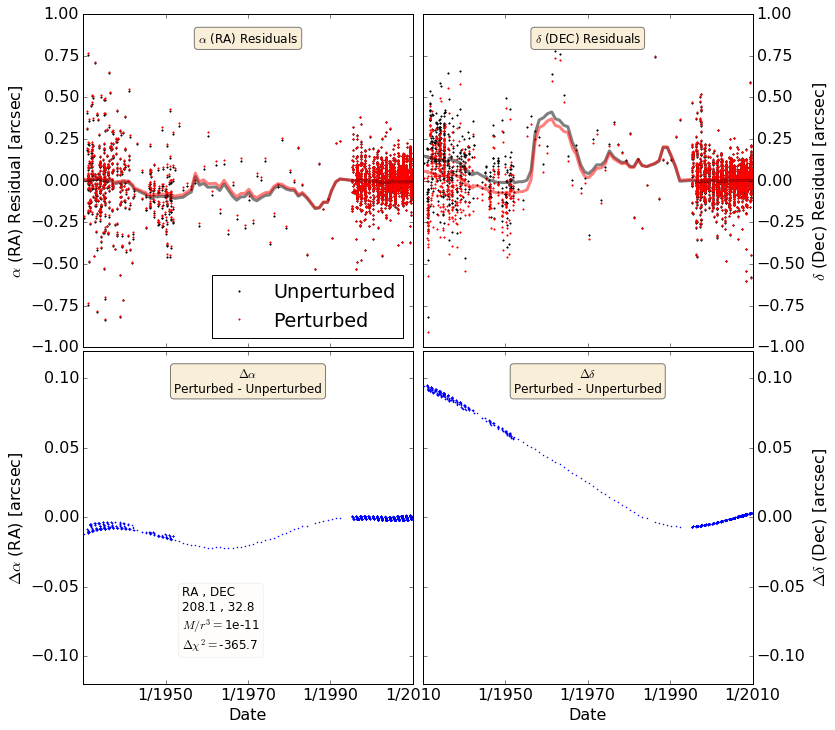}
    \caption{ %
    Residuals to orbit fits.
    {\bf Top row:} we show the RA (left) \& Dec (right) residuals to the \emph{unperturbed} orbital fit of Pluto (black) and the residuals to a \emph{perturbed} fit (red) in which the perturbing planet has sky coordinates $RA, Dec = 208.1\arcdeg, 32.8\arcdeg$ and has mass $10^{-4}M_{\odot}$ and semi-major axis $250\,AU$. 
    This corresponds to the middle plot of Fig. \ref{FIG:EXAMPLES}.
    Overplotted as transparent lines are 5-year rolling averages of the residuals for both fits, demonstrating that the data close to 1930 displays a significant, positive trend in declination residual.
    {\bf Bottom row:} we plot the difference between these models (perturbed - unperturbed). 
    It can be seen that the perturbed model (which has a $\Delta\chi^2 \sim $ -366) systematically lowers the declination residuals close to 1930, but in this example, a slight trend still remains.  
    }
    \label{FIG:RESID1}
    \end{minipage}
    \end{figure*}
    %

    \begin{figure*}
    \begin{minipage}[b]{\textwidth}
    \centering
    \includegraphics[trim = 0mm 0mm 0mm 0mm, clip, angle=0, width=\columnwidth]{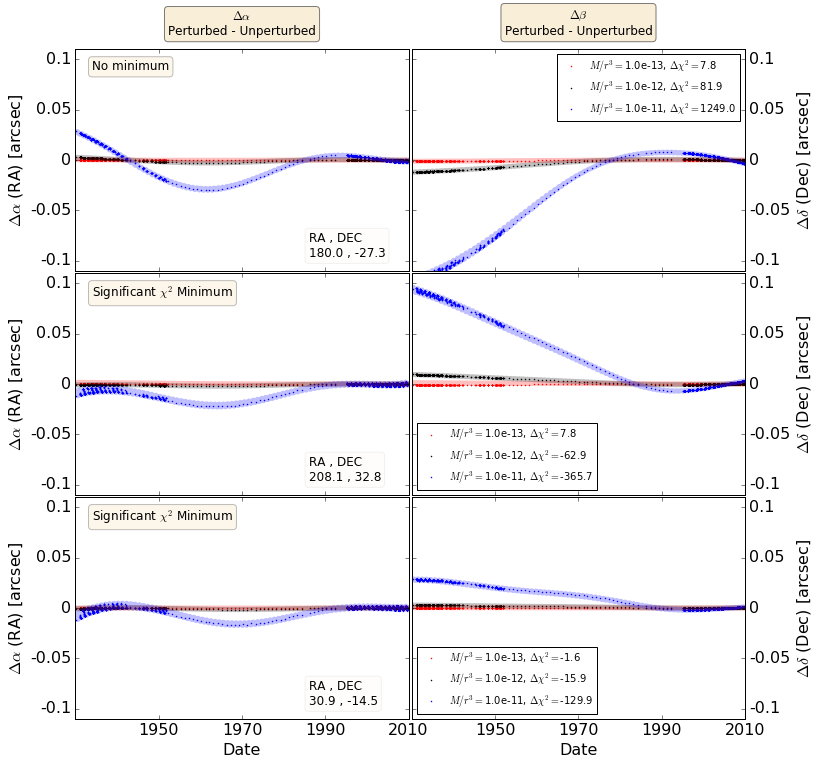}
    \caption{ %
    Residuals to orbit fits.
    {\bf Top:} we plot the RA (left) and Dec (right) differences between perturbed and unperturbed models for three different perturbation magnitudes. These models are for sky coordinates $RA, Dec = 180\arcdeg,-27\arcdeg$, corresponding to the top plot of Fig. \ref{FIG:EXAMPLES}, and a region in Fig. \ref{FIG:ALLSKY2} for which we strongly exclude perturbations down to $\sim\,10^{-12}$.
    {\bf Middle:} we plot models for sky coordinates $RA, Dec = 208.1\arcdeg, 32.8\arcdeg$. 
    This perturber position corresponds to the middle plot of Fig. \ref{FIG:EXAMPLES}, and a region in Fig. \ref{FIG:ALLSKY2} for which we find a significantly improved fit with a perturbation $\sim\,10^{-11}$.
    Note that the blue lines in the middle row are repeated from Figure \ref{FIG:RESID1}. 
    {\bf Bottom:} we repeat this analysis for sky coordinates $RA, Dec = 30.9\arcdeg, -14.5\arcdeg$, corresponding to the bottom plot of Fig. \ref{FIG:EXAMPLES} and a region in Fig. \ref{FIG:ALLSKY2} close to both the orbits of Pluto and \PX, and for which we again find a significantly improved fit with a perturbation $\sim\,10^{-11}$.
    In each of the figures, the red lines are  insignificant ( perturbation magnitudes are too small, giving $\Delta\chi^2$ close to zero).
    In the \emph{top} plot, the blue and black lines can both be excluded as the perturbation magnitudes are too large.
    In the middle and bottom plot the blue lines are close to the local $\Delta\chi^2$ minima, and it can be seen that both models cause significant modification to the Declination values close to 1930. 
    In addition to our standard model results in small, bold points, we also plot the results of the full n-body fit described in Section \ref{SECN:METHOD}, using broad, transparent lines, making it clear that there is no discernible difference between our standard method (assuming the JPL DE 432 ephemeris) and the full n-body integrations.
    }
    \label{FIG:RESID2}
    \end{minipage}
    \end{figure*}
    %

\section{Discussion}
\label{SECN:DISC}

Our results are sensitive to the tidal parameter, $M_X/r_X^3$, and insensitive to the individual mass or radius (see Figures \ref{FIG:EXAMPLES} and \ref{FIG:ALLSKY1}).
The \emph{upper limits} to the tidal parameter seen in Figures \ref{FIG:ALLSKY2} and \ref{FIG:TRUE_ANOM} span the range $10^{-12}\,-\,10^{-10}\,M_{\odot}\,AU^{-3}$, or
\begin{equation}
3 < \left(\frac{M_X}{1 M_{\oplus}}\right)\left(\frac{r_X}{100\,AU}\right)^{-3} < 300.
\end{equation}
This is very much closer and/or more massive than the nominal \PX model of \citet{Batygin.2016}, which would predict $3\times10^{-14}\,-\,10^{-12}\,M_{\odot}\,AU^{-3}$ for a $10\,M_{\oplus}$ planet in an orbit ranging from $\sim300\,-\,1,000\,AU$.

A few caveats accompany our results.  Some of these concern the data themselves and some concern our model.

\subsection{Data}

The most important caveat is that long-term systematic errors in the astrometry could strongly influence the fits.  \citet{Buie.2015} took great care to develop a reliable astrometric data set with which to better determine  Pluto's orbit.  The successful New Horizons encounter demonstrates the overall accuracy of the resulting ephemeris.  Our work is built upon that strong foundation, but systematic errors are still evident in the residuals.

In particular, a clear trend in the declination residuals can be seen in Figure~\ref{FIG:RESID1} for 1930-1950.  Although the inclusion of \PX significantly reduces this trend, it does not generally eliminate it.  We believe that the large residuals are affecting our fits, and large perturbation magnitudes (Figure \ref{FIG:RESID2}) are required to compensate for those residuals.

The early, photographic observations are the most likely to suffer from systematic errors.  The magnitude of the scatter in the astrometric measurements is understood~\citep{Buie.2015}, but the trend in declination is not. We are not aware of a systematic error that would result in a persistent trend over two decades.  It seems unlikely to be the result of zonal errors in the stellar catalogs: \citet{Buie.2015} measured Pluto's position against modern stellar catalogs, with accurate proper motion estimates.
 
The relatively few points in the 1950-1990 era, when photographic plates were still used and before CCD cameras were available, have little influence on the fits.  However, we note that the residuals in the 1960-1980 era tend to be {\it high} in declination, opposite that seen in the earlier data.  We recommend that other archival photographic plates of Pluto and other TNOs, particularly from the 1960-1980 era, be remeasured with the same care that \citet{Buie.2015} remeasured the Lampland plates.

Timing uncertainties could be substantial for the early observations, as the accuracy of the recorded observation times rely upon the care and attention of the observers, as well as the accuracy of the time-keeping itself (see ~\citealt{Buie.2015} for an interesting discussion of time-keeping at Lowell Observatory).  However, it is not clear how a trend might result from timing uncertainties. We explored allowing for time offsets for the early observations (results \emph{not} illustrated here), effectively trading one dimension of each astrometric observation for more precise determination of the observation.  Preliminarily, we find that this effectively eliminates the RA residual, but it leaves the declination residuals broadly the same.  

If the residual trend in declination were smaller it is possible that the favored perturbations would also be smaller. 
In a future publication, we explore the dependence of our results on these trends.

\subsection{Model}

The masses in our model are not fully interacting in a self-consistent manner.  However, we validated the assumptions of our model and its underlying numerical integrations (see the clear agreement in Figure~\ref{FIG:RESID2}).  Thus, the effect of ignoring some planetary interactions must be small.

The acceleration from TNOs, e.g. Haumea, can exceed that from \PX  for some brief periods of time in our integrations, but without significantly altering the results, as in almost stages of all of our integrations the dominant perturbation is from \PX.  
We estimated the masses of several of the large TNOs that we included as perturbers, and while the uncertainty in these masses obviously affects our results, any mass corrections will be small and hence remain sub-dominant to the modelled effects from \PX. 

Some features of the Kuiper belt, such as the edge near 50 AU for the cold classical, are not obviously explained by Planet Nine.  
Some investigations have argued for closer planets, possibly more that one~\citep{Brunini.2002,Lykawka.2008,BK16,KB16}.
There are undoubtedly unknown solar system masses that are not included in our model, which are yet to be discovered, and which may be significant perturbers.  There could be more than one additional massive and distant object, or there could be additional smaller and closer planets.  Investigating the influence of more than one additional planet is beyond the scope of the present investigation.  Closer planets would be more easily detected through means such as astrometric microlensing~\citep{Gaudi.2005}.

\begin{figure}[htp]
    \centering
    \includegraphics[trim = 0mm 0mm 0mm 0mm, clip, angle=0, width=\columnwidth]{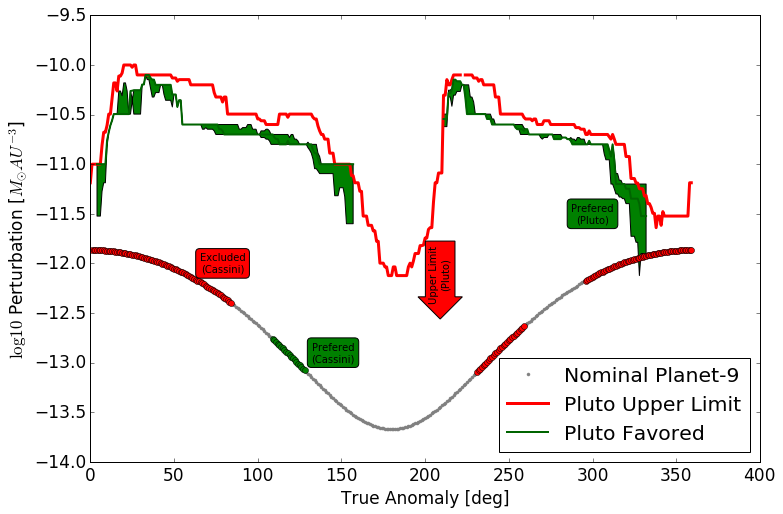}
    \caption{ Constraints on \PX arising from this work and that of \citet{Fienga.2016}. The perturbation magnitude as a function of true anomaly is plotted in grey for the nominal orbital of \PX in \citet{Batygin.2016}. The red points on the orbit are excluded by \citet{Fienga.2016}, while the green points on that line are favored. 
    For our study using observations of Pluto, our upper limits on the perturbation magnitude are plotted as the upper red line, while the perturbation magnitude range which improves the fit is plotted as a green range.
    N.B. For points on the sky \emph{away} from the nominal orbit we can place more \emph{stringent} constraints. 
    }
    \label{FIG:TRUE_ANOM}
    \end{figure}

In Figure \ref{FIG:TRUE_ANOM} we compare the results of \citet{Fienga.2016} with our own results for the specific nominal orbit of \PX~from~\citet{Batygin.2016}. 
 The tidal parameters favored by our models are larger than those suggested by \citet{Batygin.2016} and supported by the Cassini range observations~\citep{Fienga.2016}.  One possible resolution of this apparent inconsistency is the presence of more than one additional planet.  If one planet were at $60-100$~AU, closer to Pluto, it would not have to be as massive as Planet Nine to significantly perturb Pluto for a period of time.  The Cassini range observations would not necessarily be significantly affected by such a planet, because those data are only sensitive to the tidal acceleration, rather than the direct acceleration.  Furthermore, the Cassini data are from a very different range of times.

    %

\section{Conclusions}\label{SECN:CONC}

We have used astrometry of Pluto and other TNOs to constrain the sky location, distance, and mass of \PX.  We find that over broad regions of the sky the inclusion of a massive, distant planet degrades the fits to the observations.  However, in other regions, the fits are significantly improved by the addition of such a planet.  Our best fits suggest a planet that is either more massive or closer than argued for by either \citet{Batygin.2016} or \citet{Fienga.2016}.  
The trend to favor larger and closer perturbing planets is driven by the residuals to the astrometry of Pluto, remeasured from photographic plates using modern stellar catalogs~\citep{Buie.2015}, which show a clear trend in declination, over the course of two decades, that drive a preference for large perturbations.
Although this trend may be the result of systematic errors of unknown origin in the observations, a possible resolution is that the declination trend may be due to perturbations from a body, additional to \PX, that is closer to Pluto, but less massive than, \PX.


\section{Acknowledgments}

We thank Norman Murray, Gareth Williams, Cesar Fuentes, Michael Lackner, Pedro Lacerda, Gongjie Li, Joshua Winn, Scott Tremaine, Doug Finkbeiner, Avi Loeb,  Tsevi Mazeh, William Folkner, and Scott Gaudi for helpful discussions.

MJH and MJP gratefully acknowledge 
NASA Origins of Solar Systems Program grant NNX13A124G, 
NASA Origins of Solar Systems Program grant NNX10AH40G via sub-award agreement 1312645088477, 
BSF Grant Number 2012384, 
NASA Solar System Observations grant NNX16AD69G, 
as well as support from the Smithsonian 2015 CGPS/Pell Grant program.

\bibliography{references}

\end{document}